%% file: main.tex
\documentclass[10pt,twocolumn,letterpaper]{article}

\usepackage[pagenumbers]{iccv} % To force page numbers, e.g. for an arXiv version

\usepackage{bm}
\usepackage{amsmath}
\usepackage{multirow}
\usepackage{threeparttable}
\usepackage{lineno}
\usepackage{stackengine}

\definecolor{iccvblue}{rgb}{0.21,0.49,0.74}
\usepackage[pagebackref,breaklinks,colorlinks,allcolors=iccvblue]{hyperref}

\def\paperID{13995} % *** Enter the Paper ID here
\def\confName{ICCV}
\def\confYear{2025}

\title{Learned Image Compression with Hierarchical Progressive Context Modeling}

\author{
Yuqi Li\thanks{
    Equal contribution. This work is supported by the Natural Science Foundation of China under Grant 62021001. We acknowledge the support of GPU cluster built by MCC Lab of Information Science and Technology Institution, USTC. (Corresponding author: Dong Liu.)
} \quad 
Haotian Zhang\textsuperscript{*} \quad 
Li Li \quad 
Dong Liu \\
{\small MOE Key Laboratory of Brain-Inspired Intelligent Perception and Cognition}\\
{\small University of Science and Technology of China, Hefei 230093, China}\\
{\tt\small \{\href{mailto:lyq010303@mail.ustc.edu.cn}{lyq010303}, \href{mailto:zhanghaotian@mail.ustc.edu.cn}{zhanghaotian}\}@mail.ustc.edu.cn, 
\{\href{mailto:lil1@ustc.edu.cn}{lil1}, \href{mailto:dongeliu@ustc.edu.cn}{dongeliu}\}@ustc.edu.cn}
}

\begin{document}
\maketitle
\input{sec/0_abstract}    
\input{sec/1_intro}
\input{sec/2_related_works}
\input{sec/3_method}
\input{sec/4_experiments}
\input{sec/5_conclusion}
{
    \small
    \bibliographystyle{ieeenat_fullname}
    \bibliography{main}
}

% \maketitle
\clearpage
\renewcommand{\thetable}{\Alph{table}}
\renewcommand{\thefigure}{\Alph{figure}}
\setcounter{figure}{0}
\setcounter{table}{0}
% \setcounter{page}{1}
% \maketitlesupplementary
\twocolumn[{%
\renewcommand\twocolumn[1][]{#1}%
\maketitlesupplementary
\begin{center}
    \centering
    \includegraphics[width=1.0\textwidth]{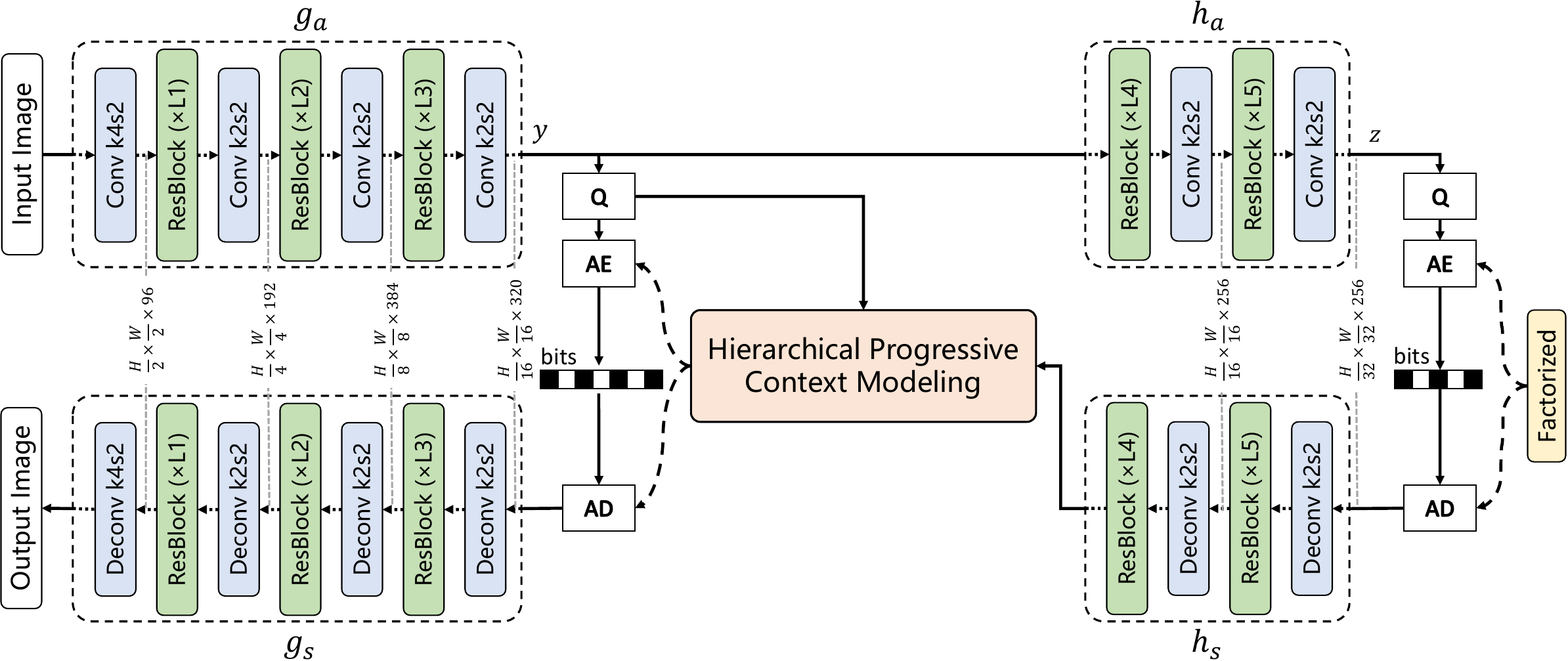}
    \captionof{figure}{Overall architecture of our proposed model. The detailed structure of the ResBlock is shown in Fig.~\ref{blocks}. `k2s2' represents the convolution layer with kernel size as 2 and stride as 2.}
    \label{arch}
\end{center}%
}]

\setcounter{section}{0}
\renewcommand\thesection{\Alph{section}}
\section{Overall architecture}
\label{overall_arch}

The overall architecture of our proposed model is shown in Fig.~\ref{arch}. Following the design of transform networks in previous learned image compression methods \cite{he2022elic}, we adopt residual blocks along with down-sampling and up-sampling layers to establish nonlinear transforms $g_a$, $g_s$, $h_a$ and $h_s$.
Specifically, we use a single convolution layer with a stride of 2 or 4 for both down-sampling and up-sampling operations. The kernel size of the input down-sampling layer and output up-sampling layer is set as 4, and the kernel sizes in other down-sampling layers and up-sampling layers are set as 2 to reduce complexity.
Additionally, we utilize the advanced FasterNet \cite{chen2023pconv} blocks as the ResBlock in our model. The detailed structure of the ResBlock is presented in Fig.~\ref{blocks} (a). The PConv is the partial convolution layer, which processes spatial dense convolution only on partial channels.
In our HPCM-Base model, the number of ResBlocks at different stages is set as $[L1, L2, L3, L4, L5] = [2, 2, 4, 1, 3]$. In our HPCM-Large model, the depth of $L3$ stage is increased for enhanced transform capacity, $[L1, L2, L3, L4, L5]=[2, 2, 8, 1, 3]$.

\section{Structure of Entropy Parameter Networks}
\label{structure_entropy}

Figure \ref {gep} shows the network structure of the entropy parameter network $g_{ep}$. $g_{ep}$ contains a 1×1 convolution layer followed by multiple DepthConvBlocks. The detailed structure of the DepthConvBlock is shown in Fig.\ref{blocks} (b).
To reduce the model parameters, we share the weights of DepthConvBlocks across different coding steps. 
We use two different entropy parameter networks. One is used for coding $\hat{y}^{S1}$ and $\hat{y}^{S2}$, denoted as $g_{ep}^{S1+S2}$. 
Another is used for coding $\hat{y}^{S3}$, denoted as $g_{ep}^{S3}$.
% Specifically, the same DepthConvBlocks are used for coding , denoted as $DW_1$, and for coding $\hat{y}^{S3}$, denoted as $DW_2$. 
To enhance the adaptability of shared networks, we introduce step adaptive embedding into entropy parameter networks to modulate the weight of each channel.
In our HPCM-Base model, the numbers of DepthConvBlocks are set as $[N1, N2] = [2, 1]$ and $[N1, N2] = [3, 2]$ in $g_{ep}^{S1+S2}$ and $g_{ep}^{S3}$, respectively. In our HPCM-Large model, the numbers of DepthConvBlocks are set as $[N1, N2] = [2, 2]$ and $[N1, N2] = [4, 3]$ in $g_{ep}^{S1+S2}$ and $g_{ep}^{S3}$, respectively. 
% we increase the depth, setting $DW_1$ to $[N1, N2] = [2, 2]$ and $DW_2$ to $[N1, N2] = [4, 3]$.
% Except for using the synthesized hyperprior to obtain the entropy parameter state $\psi_{i}$ at the first coding step, we derive $\psi_{i}$ using $g_{ep}$. At the $i_{th}$ coding step, the input to $g_{ep}$ is is the concatenation of $\hat{y}_{<i}$ and $C_i$.

\begin{figure*}[!t]
    \centering
    \centerline{
        \begin{minipage}{0.45\linewidth}
         \includegraphics[width=\textwidth]{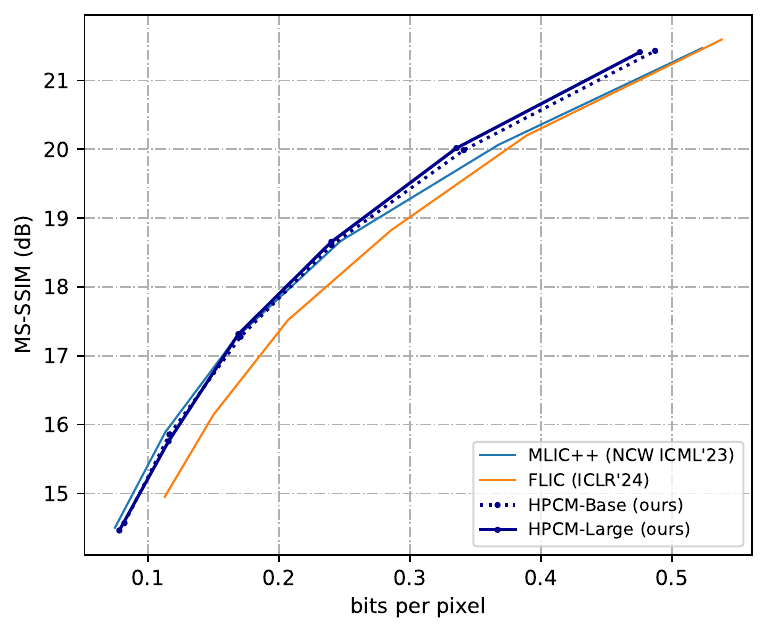}
         % \vspace{0.0001cm}
        \end{minipage}
        \hspace{1cm}
        \begin{minipage}{0.45\linewidth}
         \includegraphics[width=\textwidth]{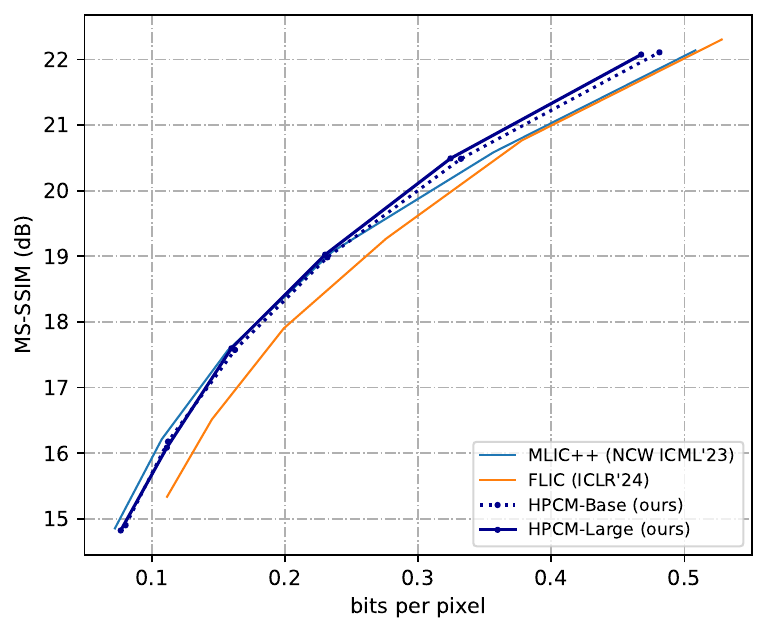}
         % \vspace{0.0001cm}
        \end{minipage}
    }
    \vspace{-1em}
    \caption{Rate-distortion curves on different datasets. The left one is tested on the CLIC Pro Valid dataset, and the right one is tested on the Tecnick dataset.}
    \label{rd}
\end{figure*}

\begin{figure}[!t]
    \centering
    \centerline{\includegraphics[width=0.4\textwidth]{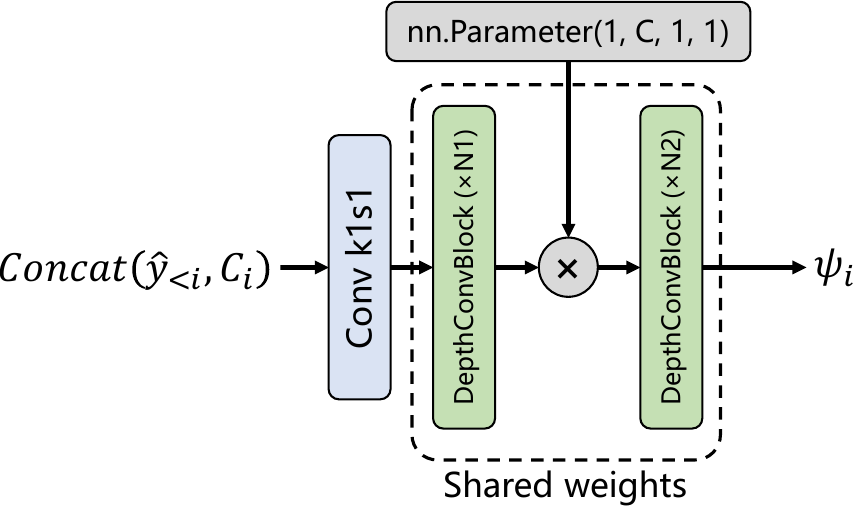}}
    \caption{The network structure of entropy parameter network $g_{ep}$. The detailed structure of the DepthConvBlock is illustrated in Fig.~\ref{blocks}.}
    \label{gep}
\end{figure}

\begin{figure}[!t]
    \centering
    \centerline{\includegraphics[width=0.4\textwidth]{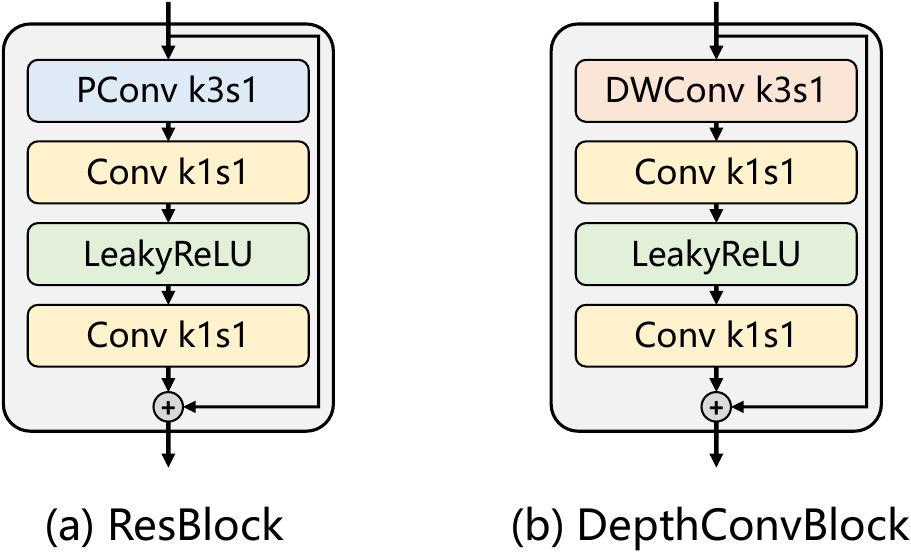}}
    \caption{The detailed structure of the ResBlock and DepthConvBlock.}
    \label{blocks}
\end{figure}

% \begin{figure}[!t]
%     \centering
%     \centerline{\includegraphics[width=0.45\textwidth]{figs/rd_clic_ssim.pdf}}
%     \caption{Rate-distortion curves on CLIC Pro Valid dataset.}
%     \label{rd_clic}
% \end{figure}

% \begin{figure}[!t]
%     \centering
%     \centerline{\includegraphics[width=0.45\textwidth]{figs/rd_tec_ssim.pdf}}
%     \caption{Rate-distortion curves on Tecnick dataset.}
%     \label{rd_tecnick}
% \end{figure}

\section{Detailed Hierarchical Coding Schedule}
\label{detailed_coding}

\begin{figure*}[!t]
    \centering
    \centerline{\includegraphics[width=1.0\textwidth]{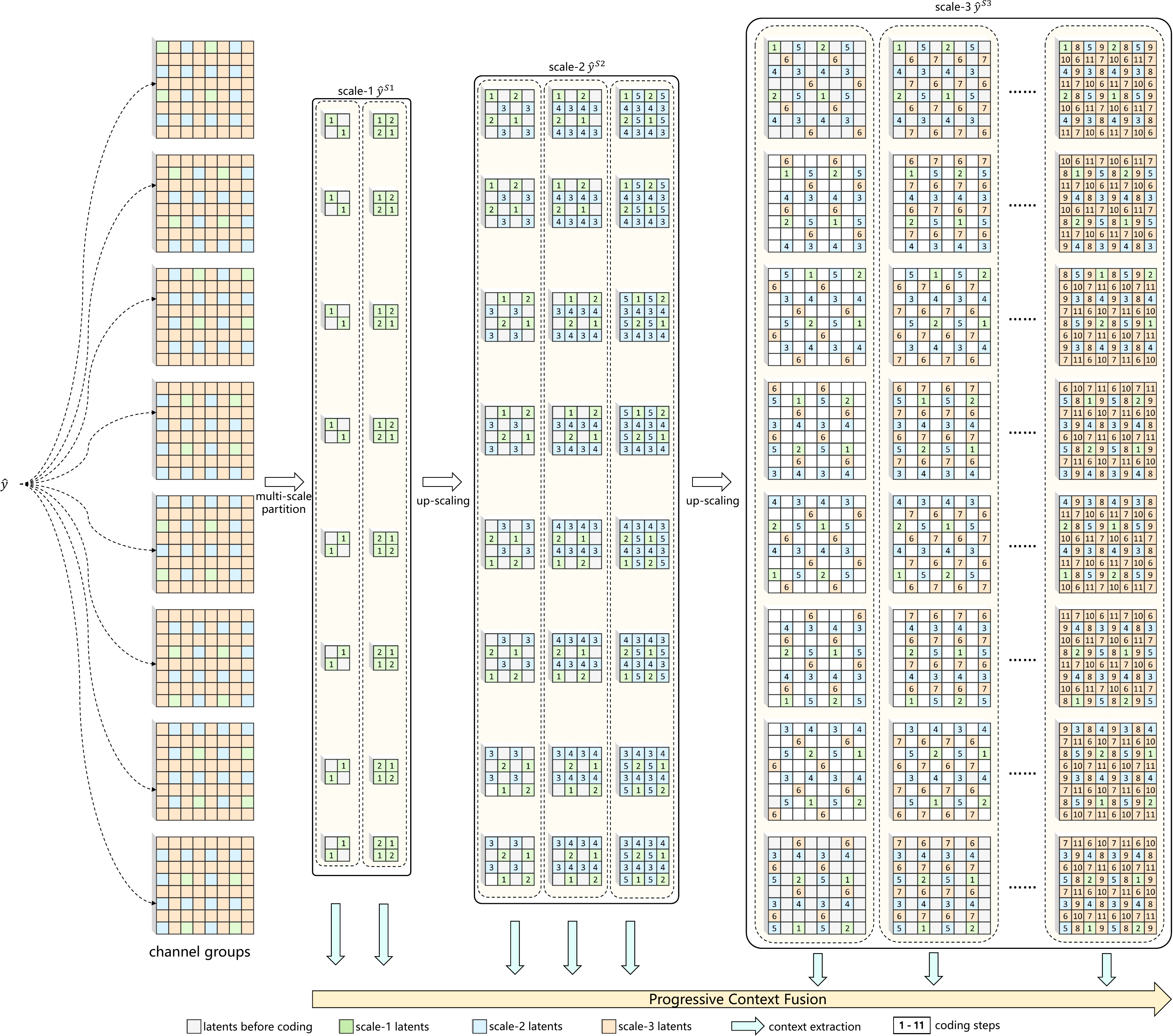}}
    \caption{The detailed coding progress of our proposed hierarchical coding schedule.}
    \label{order}
\end{figure*}

Figure \ref{order} presents the detailed multi-scale partition process and coding schedules on eight channel groups of $\hat{y}$. Different multi-scale partition methods are applied to different channel groups to enable the interaction of spatial and channel-wise context information. We design the coding schedule on $\hat{y}^{S3}$ following previous studies \cite{li2023flexible, mentzer2023m2t}, which fully exploits the spatial correlation and enhances context diversity.

% \section{Coding Step Allocation of Hierarchical Coding Schedule}

\begin{figure*}[!t]
    \centering
    \centerline{\includegraphics[width=0.97\textwidth]{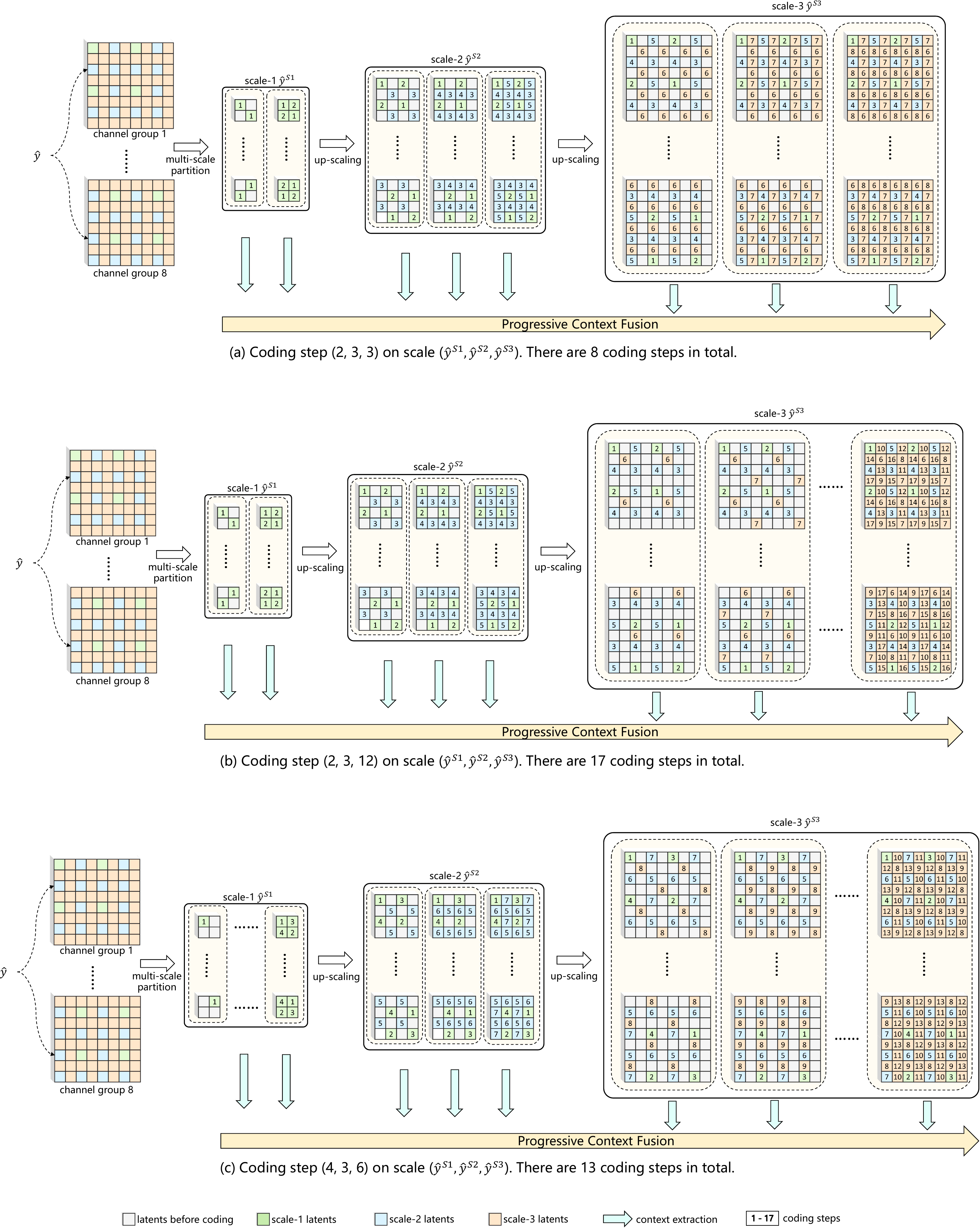}}
    \caption{The detailed coding process for (a) coding step (2, 3, 3), (b) coding step (2, 3, 12), and (c) coding step (4, 3, 6) for coding $(\hat{y}^{S1}, \hat{y}^{S2}, \hat{y}^{S3})$.}
    \label{order_abl}
\end{figure*}

Figure \ref{order_abl} illustrates the detailed coding process for different coding step allocations. We present the coding step (2, 3, 3), (2, 3, 12), and (4, 3, 6) for coding $(\hat{y}^{S1}, \hat{y}^{S2}, \hat{y}^{S3})$, respectively.

\begin{table}[!t]
\centering
\caption{Comparison of training speed of various methods.}
\begin{tabular}{lc}
\hline
Model                & \begin{tabular}[c]{@{}c@{}}Training Speed $\uparrow$ \\ (steps/s)\end{tabular} \\ \hline
ELIC (CVPR'22) \cite{he2022elic}       & 4.07                                                                \\
STF (CVPR'22) \cite{zou2022devil}        & 3.31                                                                \\
TCM (CVPR'23) \cite{liu2023learned}        & 1.28                                                                \\
MLIC++ (NCW ICML'23) \cite{jiang2023mlic} & 1.24                                                                \\
FLIC (ICLR'24) \cite{li2023frequency}       & 1.93                                                                \\
WeConvene (ECCV'24) \cite{fu2024weconv}  & 1.43                                                                \\ \hline
HPCM-Base (ours)     & 3.67                                                                \\
HPCM-Large (ours)    & 2.88                                                                \\ \hline
\end{tabular}
\label{train_time}
\end{table}

\section{Comparison of Training Speed}

We compare the training speed of the proposed HPCM-Base and HPCM-Large with recent advanced methods in Table~\ref{train_time}. The training time is evaluated on one NVIDIA GeForce RTX 3090 GPU, with training batch size as 8 images and patch size as 256×256. Compared to recent state-of-the-art learned image compression methods like \cite{jiang2023mlic,li2023frequency}, our methods achieve faster training speed.

\section{Additional Rate-Distortion Performance}

Figure~\ref{rd} presents the rate-distortion performance of our model optimized for MS-SSIM.
% on CLIC Pro Valid and Tecnick datasets. We use MS-SSIM to represent the distortion term. 
Compared to recent advanced learned image compression methods, our proposed methods achieve superior performance at higher bitrates.

\begin{table*}[!t]
\centering
\caption{Code links of various methods. We use these open-source implementations to evaluate the compression performance and computational complexity of each model.}
\begin{tabular}{ll}
\hline
Model                                     & Code Link                                             \\ \hline
ELIC (CVPR'22) \cite{he2022elic}          & \url{https://github.com/JiangWeibeta/ELIC}            \\
STF (CVPR'22) \cite{zou2022devil}         & \url{https://github.com/Googolxx/STF}                 \\
TCM (CVPR'23) \cite{liu2023learned}       & \url{https://github.com/jmliu206/LIC_TCM}             \\
MLIC++ (NCW ICML'23) \cite{jiang2023mlic} & \url{https://github.com/JiangWeibeta/ELIC}            \\
FLIC (ICLR'24) \cite{li2023frequency}     & \url{https://github.com/qingshi9974/ICLR2024-FTIC}    \\
MambaVC (Arxiv'24) \cite{qin2024mambavc}  & \url{https://github.com/QinSY123/2024-MambaVC}        \\
WeConvene (ECCV'24) \cite{fu2024weconv}   & \url{https://github.com/fengyurenpingsheng/WeConvene} \\ \hline
\end{tabular}
\label{code_link}
\end{table*}

\begin{table}[!t]
\centering
% \vspace{-0.8em}
\fontsize{9pt}{12}\selectfont
\setlength{\tabcolsep}{1.1pt}
% \renewcommand{\arraystretch}{0.85}
% \vspace{-1.0em}
\caption{Coding times of various methods. Times are in milliseconds (ms).}
\begin{tabular}{c|ccc|ccc|c}
\hline
\multirow{2}{*}{\Centerstack[c]{Models}} & \multicolumn{3}{c|}{\Centerstack[c]{Encoding}} & \multicolumn{3}{c|}{\Centerstack[c]{Decoding}}& \multirow{2}{*}{\Centerstack[c]{kMACs/pixel}} \\
& $T_{Net}$ & $T_{AC}$ & $T_{Total}$ & $T_{Net}$ & $T_{AC}$ & $T_{Total}$ &  \\ \hline
ELIC                                                    & 44        & 82       & 126       & 37        & 74       & 111       & 573.88                                                    \\
TCM                                                     & 89        & 110      & 199       & 83        & 119      & 202       & 1823.58                                                   \\
MLIC++                                                  & 73        & 117      & 190       & 74        & 152      & 226       & 1282.81                                                   \\ \hline
% \begin{tabular}[c]{@{}c@{}}ELIC\\ (our AC)\end{tabular} & 50        & 23       & 73        & 34        & 45       & 79        & 574                                                    \\
HPCM-Base                                               & 58        & 25       & 83        & 57        & 24       & 81        & 918.57                                                    \\ \hline
\end{tabular}
% \vspace{-1.0em}
\label{TIME}
% \vspace{-2.4em}
\end{table}

\begin{table}[htbp]
\centering
% \vspace{-1.15em}
% \fontsize{8pt}{10}\selectfont
% \setlength{\tabcolsep}{1.5pt}
% \renewcommand{\arraystretch}{0.85}
% \renewcommand{\thetable}{B}
% \vspace{-0.95em}
\caption{Abaltion studies on the number of hierarchical stages.}
\begin{tabular}{ccc}
\hline
Models              & kMACs/pixel & BD-Rate \\ \hline
w/o hierarchical (1-stage)             & 1107.48     & 1.07\%  \\
2-stage             & 954.17      & 0.35\%  \\
HPCM-Base (3-stage) & 918.57      & \textbf{0.00\%}  \\
4-stage             & \textbf{911.44}      & 0.18\%  \\ \hline
\end{tabular}
% \vspace{-1.0em}
\label{stage}
% \vspace{-1.4em}
\end{table}

\section{Additional Results on Coding Time}
\label{coding_time}

Table~\ref{TIME} breaks down coding times into network inference ($T_{Net}$) and arithmetic coding ($T_{AC}$).
For prior studies, we used their open-sourced models and code, as shown in Table~\ref{code_link}.
Across different models, $T_{Net}$ generally increases with higher kMACs/pixel; this aligns with HPCM-Base exhibiting a higher $T_{Net}$ than ELIC. 
As for $T_{AC}$, it varies across models due to different implementations in the released code. 
Our arithmetic coding implementation improves upon the widely used CompressAI-based \cite{begaint2020compressai} implementation by enabling more efficient data exchange between Python and C.
This optimization significantly reduces $T_{AC}$.

Since coding time is implementation-dependent, we focus on comparing the kMACs/pixel to measure computational complexity. Fig. \ref{fig1} in the maintext plots ``kMACs/pixel vs. BD-Rate", showing that our method achieves superior performance-complexity trade-offs compared to current SOTA methods.

\section{Additional Ablation Studies}
\subsection{Ablation Studies on Hierarchical Coding Stages}
\label{add_abla_hie}

% By keeping the total number of coding steps fixed for a fair comparison, we allocate some coding steps to lower latent scales when introducing additional stages. Context extraction at lower scales is more efficient compared to higher scales due to smaller input sizes, thereby reducing overall complexity.

We further provide ablation studies to verify the superiority of the 3-stage model. As shown in Table~\ref{stage}, the 3-stage model achieves lower kMACs/pixel and better performance compared to the 1-stage and 2-stage variants, as it more effectively captures long-range spatial contexts.
\begin{table}[t]
\centering
% \vspace{-1.1em}
% \fontsize{8pt}{10}\selectfont
\setlength{\tabcolsep}{1.5pt}
% \renewcommand{\arraystretch}{0.9}
% \renewcommand{\thetable}{C}
% \vspace{-1.1em}
\caption{Ablation studies on shared parameters in context models.}
\begin{tabular}{cccc}
\hline
Models           & kMACs/pixel & Params (M) & BD-Rate \\ \hline
HPCM-Base         & 918.57      & 68.50  & 0.00\%  \\
w/o shared params & 918.57      & 189.99 & -0.15\%  \\ \hline
\end{tabular}
% \vspace{-1.0em}
\label{share_nets}
% \vspace{-2.2em}
\end{table}
Increasing the number of stages to 4 slightly reduces computational complexity, but the performance is marginally worse. This is because allocating coding steps to capture extremely long-range spatial contexts is not cost-effective. Therefore, we adopt the 3-stage model as HPCM-base.

\subsection{Ablation Studies on Shared Parameters of Context Models}

We have tested a variant of HPCM-Base using context models with non-shared parameters at \textbf{all} scales. As shown in Table~\ref{share_nets}, this model achieves comparable performance but with a substantial increase in the number of parameters. This indicates that employing different network settings across scales offers limited benefits to our model. Therefore, we shared the parameters of context models at all scales to significantly reduce the model's parameter count.

\section{Additional Visual Comparison Results}

\begin{figure*}[!t]
    \centering
    \centerline{\includegraphics[width=1.0\textwidth]{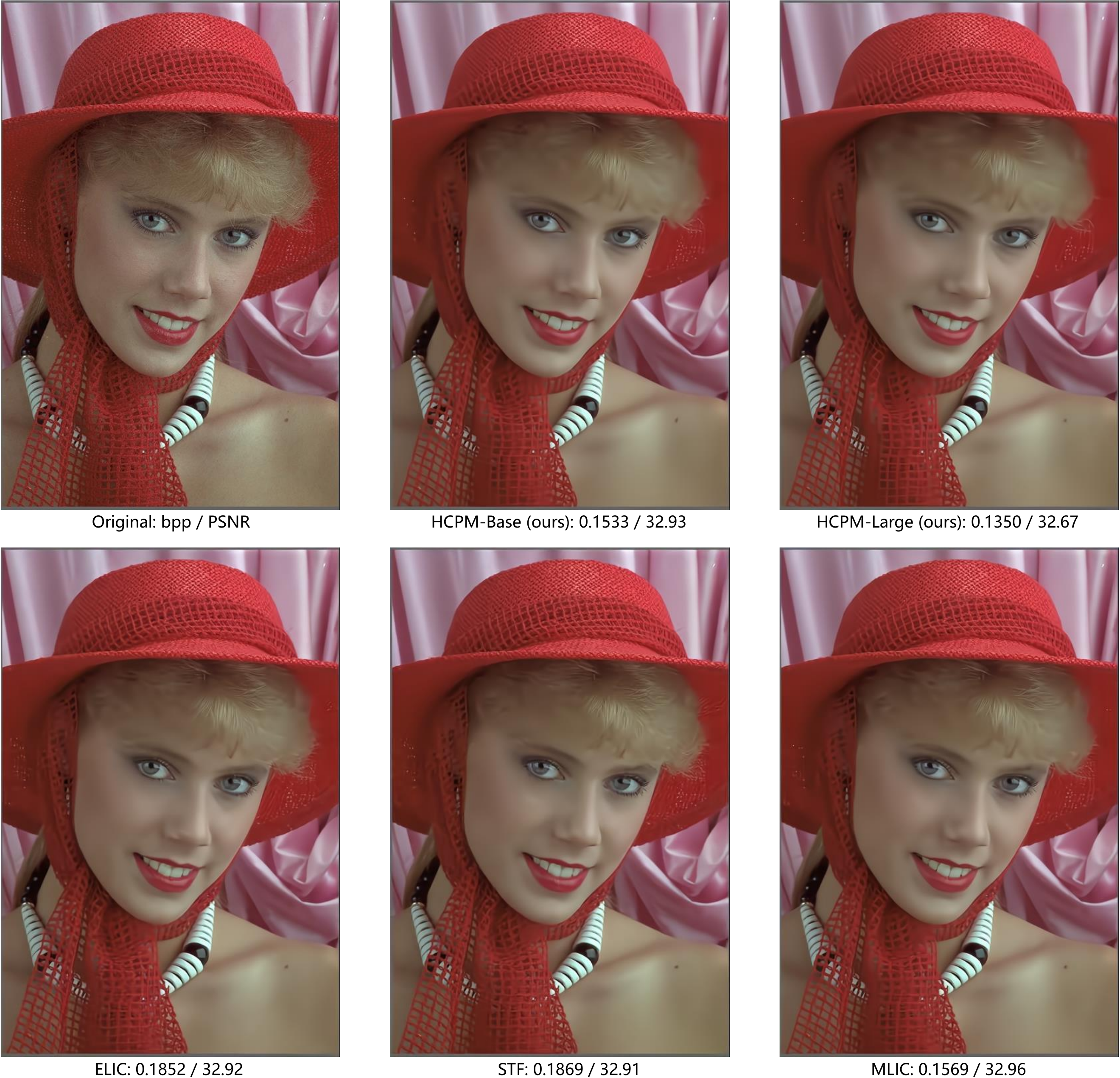}}
    \caption{Visual comparison of reconstructed images of Kodim04 in the Kodak dataset with various learned image compression methods.}
    \label{v1}
\end{figure*}

\begin{figure*}[!t]
    \centering
    \centerline{\includegraphics[width=1.0\textwidth]{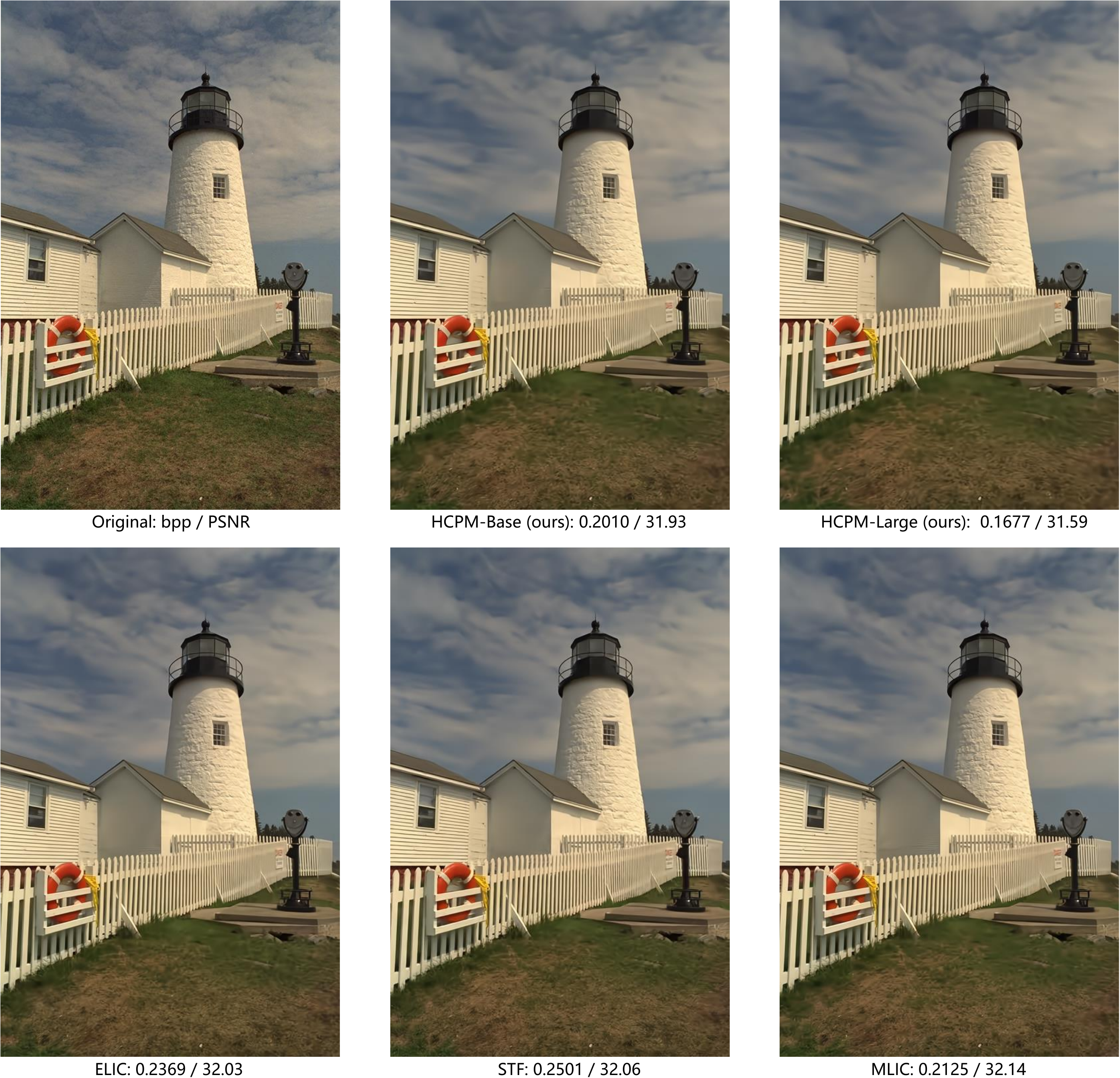}}
    \caption{Visual comparison of reconstructed images of Kodim19 in the Kodak dataset with various learned image compression methods.}
    \label{v2}
\end{figure*}

Fig.~\ref{v1} and Fig.~\ref{v2} presents the reconstructed images of our proposed HPCM-Base, HPCM-Large, and various methods \cite{he2022elic, zou2022devil, jiang2023mlic}.

% {
%     \small
%     \bibliographystyle{ieeenat_fullname}
%     \bibliography{main}
% }

\end{document}

%% file: sec/0_abstract.tex
\begin{abstract}
Context modeling is essential in learned image compression for accurately estimating the distribution of latents. While recent advanced methods have expanded context modeling capacity, they still struggle to efficiently exploit long-range dependency and diverse context information across different coding steps.
In this paper, we introduce a novel Hierarchical Progressive Context Model (HPCM) for more efficient context information acquisition.
Specifically, HPCM employs a hierarchical coding schedule to sequentially model the contextual dependencies among latents at multiple scales, which enables more efficient long-range context modeling.
Furthermore, we propose a progressive context fusion mechanism that incorporates contextual information from previous coding steps into the current step, effectively exploiting diverse contextual information.
Experimental results demonstrate that our method achieves state-of-the-art rate-distortion performance and strikes a better balance between compression performance and computational complexity. The code is available at \url{https://github.com/lyq133/LIC-HPCM}.
% 先建一个空仓库吧
\end{abstract}

%% file: sec/1_intro.tex
% \vspace{-1em}
\section{Introduction}
\label{sec:intro}
Image compression is a fundamental topic in the field of computer vision and image processing, which plays a crucial role in enabling efficient visual data storage and transmission. In the passing decades, many traditional codecs have achieved impressive compression performance with manual design for each module, including JPEG \cite{wallace1991jpeg}, JPEG2000 \cite{skodras2001jpeg}, BPG \cite{Bellard2015BPG}, and VVC \cite{bross2021overview}.

In recent years, various learned compression methods \cite{he2022elic, liu2023learned, jiang2023mlic, li2023frequency} have demonstrated superior performance. 
Most learned image compression models are based on the transform coding scheme \cite{goyal2001transform}.
In such a model, the image is transformed to produce quantized latent representations. Subsequently, the quantized latents are entropy-coded into bitstreams through an entropy model. The decoder recovers the quantized latents and then applies a synthesis transform to reconstruct the image. 
The entropy model is used to estimate the distribution of latent variables for entropy coding. A more accurate entropy model leads to fewer bits, thereby improving compression performance.

\begin{figure}[!t]
  \centering
    \includegraphics[width=0.98\linewidth]{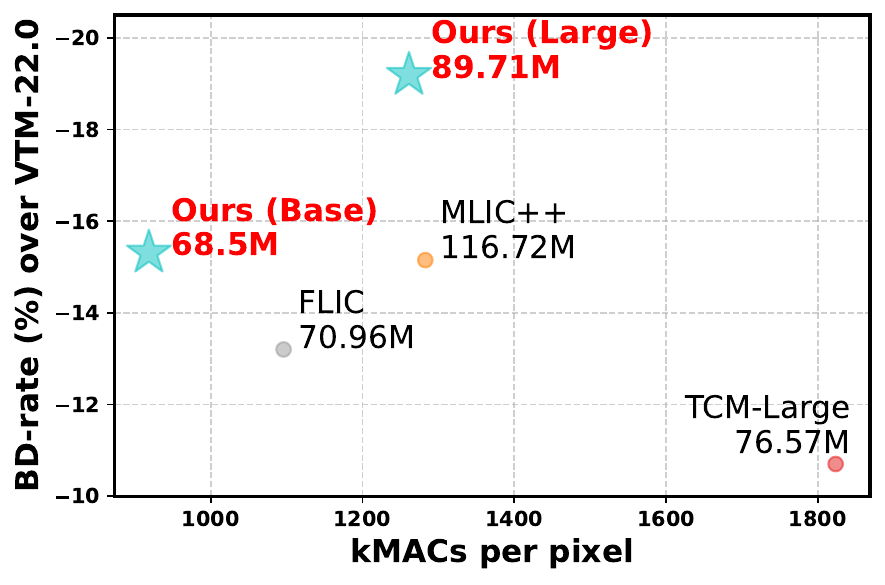}
    
    \caption{Comparison between BD-rate over VTM-22.0 and end-to-end computational complexity on Kodak dataset. We select recent advanced learned image compression methods \cite{liu2023learned, jiang2023mlic, li2023frequency} for comparison. Model parameters of various methods are noted in this figure. Top-left is better.
    }
    \label{fig1}
    \vspace{-0.5em}
\end{figure}

Most advanced learned image compression models apply conditional entropy models for entropy coding, which jointly utilize the hyperprior model \cite{balle2018variational} and context models \cite{minnen2018joint, minnen2020channel}. The context model divides latents into several groups and utilizes sequential autoregression to capture contextual information, thereby enabling accurate distribution estimation. Some studies utilized spatial context modeling \cite{minnen2018joint, he2021checkerboard} to exploit the local spatial redundancy in the latent domain. Recent studies \cite{qian2022entroformer} have explored long-range spatial context modeling with transformer architectures, while introducing higher complexity. Some studies \cite{he2021checkerboard, li2023neural} exploited context information in both channel and spatial dimensions. Jiang \textit{et al.} \cite{jiang2023mlic} introduced a multi-reference context model to effectively capture local spatial, global spatial, and channel-wise contextual information. While expanding the diversity of contextual information can boost compression performance, existing approaches still struggle to efficiently exploit long-range dependency and diverse context across different coding steps. Therefore, a more efficient method for exploiting diverse contexts is required to achieve a better trade-off between compression performance and computational complexity.

In this paper, we introduce a Hierarchical Progressive Context Model (HPCM) for more efficient context information acquisition. Our HPCM employs a hierarchical coding schedule to sequentially model contextual dependencies in latents at multiple scales. First, we partition the latents into multi-scale sub-latents through a specialized sampling method. 
Then, we sequentially code each scale of sub-latents, starting from the smallest scale and progressing to the largest, gradually modeling dependencies from long-range to short-range. 
Capturing context representations at a smaller scale of sub-latents enables more efficient modeling of long-range contextual dependencies. 
Furthermore, to better exploit the diverse context information at different coding steps, we present a progressive context fusion mechanism. Specifically, we incorporate the context information from previous coding steps into the current step's contextual representations through a cross-attention mechanism. This approach enables the progressive accumulation of diverse context information. Equipped with this progressive context fusion, our HPCM enables the efficient combination of diverse contexts at different coding steps.

% To achieve a better balance between performance and complexity, we utilize advanced neural networks \cite{chen2023pconv,chollet2017xception} to construct transforms and context model networks. 
% Additionally, we present a parameter-efficient way to construct our HPCM, which shares the same network at different coding steps. 
% To achieve a better balance between performance and complexity, we present some model improvements on network structures \cite{chen2023pconv,chollet2017xception}, parameter-efficient context model and arithmetic coder optimization. 
To better balance compression performance and complexity, we introduce structural improvements and a parameter-efficient context model design, and optimize the arithmetic coding implementation.
With the help of the proposed HPCM and model improvement, our method achieves state-of-the-art rate-distortion performance.
As shown in Fig. \ref{fig1}, compared to VTM-22.0, our Base model and Large model achieve 15.31\% and 19.19\% rate savings on the Kodak dataset, respectively. Meanwhile, our methods strike a better balance between compression performance and computational complexity. 
Our contributions can be summarized as follows: 
\begin{itemize}
    \item We present a hierarchical coding schedule for learned image compression, gradually modeling dependencies in the latent domain from long-range to short-range. This enables more efficient long-range context modeling.
    \item We propose a progressive context fusion mechanism to enable the accumulation of diverse context, which incorporates the context from previous coding steps into the current step's contextual representations.
    \item Our method achieves state-of-the-art compression performance and strikes a better balance between compression performance and complexity.
    % \vspace{0.5em}
\end{itemize}

%% file: sec/2_related_works.tex
\vspace{-1em}
\section{Related Work}
\label{sec:formatting}
\subsection{Learned Image Compression}
In recent years, learned compression methods \cite{li2024ustc, li2024uniformly, li2024object, li2024loop, tang2025neural, li2021deep, li2023neural, li2024neural, jia2025towards, hu2021fvc} have demonstrated impressive performance. Through the end-to-end optimization techniques and deep neural networks, learned image compression methods \cite{he2022elic, liu2023learned, jiang2023mlic, li2023frequency} have achieved impressive performance. Most recent studies follow the joint rate-distortion optimization scheme \cite{balle2017end} for advanced compression performance. 

The capacity of transform is important for learned image compression. A convolution network-based transform is proposed in \cite{balle2017end}. Subsequent studies \cite{chen2021end,cheng2020learned,zou2022devil,he2022elic} utilized more powerful network structures to enhance transforms, such as deep residual connections and non-local attention mechanisms. The transformer-based transforms \cite{zhu2022transformer,liu2023learned,li2023frequency} are constructed to enable the global modeling ability. Invertible neural networks have also been explored in \cite{ma2020end}.

The entropy model is used to estimate the probabilistic distribution of the latents. 
In the conditional entropy model, the distribution of latents is modeled as a specific probability distribution family, such as the Gaussian model and generalized Gaussian model \cite{zhang2024ggm}. The hyperprior and context models are used to estimate the entropy parameters. For example, in \cite{balle2018variational}, hyperprior is used for estimating the Gaussian scale parameters for characterizing the distribution of latents. Numerous studies have contributed to developing the context models, which are introduced in Sec. \ref{sec:context_model}.

Some studies enhance the performance of learned image compression models through advancing the quantization strategies \cite{zhang2023uniform, guo2021soft}. Some studies aims to explore practical learned image compression models \cite{wang2024asym, li2024deviation, zhang2025learning}.

\begin{figure*}[!t]
    \centering
    \centerline{\includegraphics[width=0.99\textwidth]{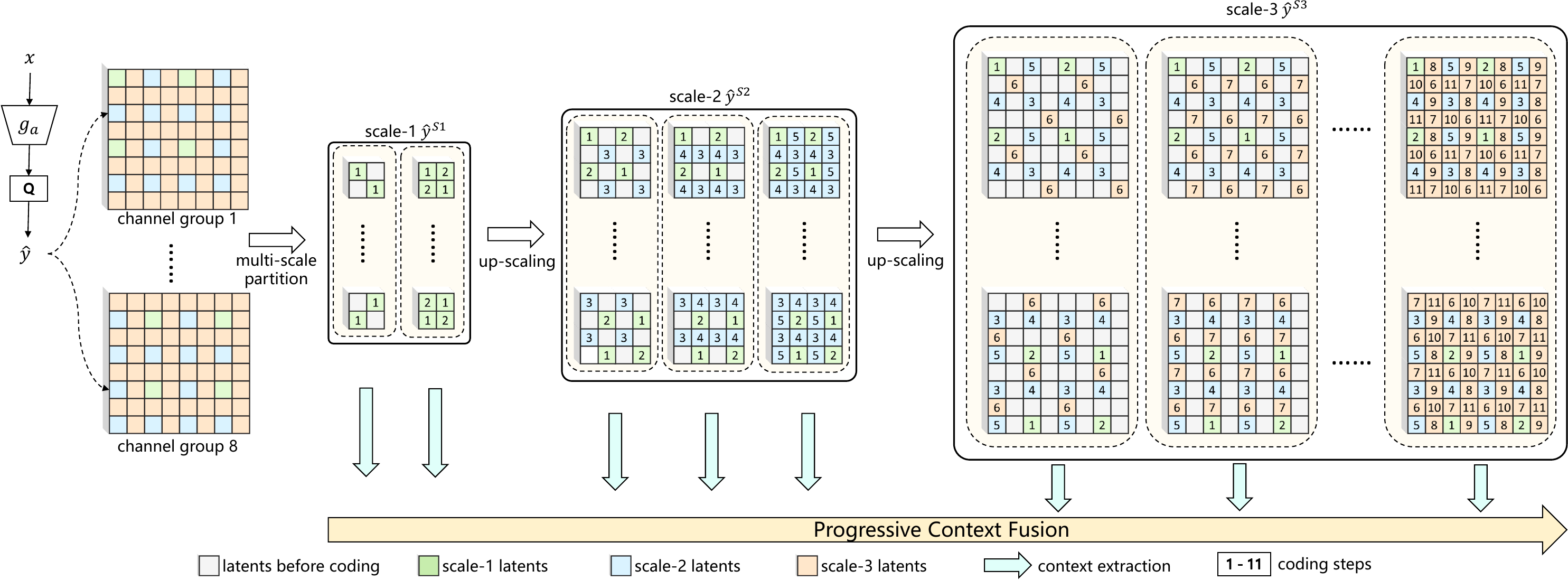}}
    \vspace{-0.5em}
    \caption{
    Hierarchical coding schedule of HPCM. 
    $\hat{y}$ is partitioned into three sub-latents at different scales: $\hat{y}^{S1}$, $\hat{y}^{S2}$, and $\hat{y}^{S3}$. 
    These partition strategies vary across different channel groups. We show the partition process of the first and eighth channel groups in this figure. 
    Sequential conditional entropy coding and progressive context modeling are then applied at each coding step. 
    We begin the coding process on $\hat{y}^{S1}$, capturing long-range dependency first. Once coding $\hat{y}^{S1}$, we fill $\hat{y}^{S1}$ back to the corresponding positions in $\hat{y}^{S2}$ for up-scaling. Similarly, after coding $\hat{y}^{S2}$, we fill $\hat{y}^{S2}$ into $\hat{y}^{S3}$. The coding steps at each scale are allocated based on the number of latent elements in each scale. The coding step for each element is noted in this figure. There are 11 coding steps in total.
    }
    \vspace{-0.5em}
    \label{overall_architecture}
\end{figure*}

% \vspace{-0.25em}
\subsection{Context Models}
\label{sec:context_model}
% \vspace{-0.25em}
The context model is effective for exploring the redundancies in latents. Specifically, it divides latents into several groups and utilizes sequential autoregression to capture contextual information, thereby enabling accurate distribution estimation.
The raster-scan autoregressive spatial context model \cite{minnen2018joint} is first adopted to exploit the local spatial redundancy in latents. Qian \textit{et al.} \cite{qian2022entroformer} further explored long-range spatial context modeling with transformer architectures, while introducing higher complexity.
He \textit{et al.} \cite{he2021checkerboard} proposed a checkerboard  model for parallel spatial context modeling. From another perspective, Minnen \textit{et al.} \cite{minnen2020channel} proposed an efficient channel-wise context model. Some studies \cite{he2022elic,li2023neural} exploited context information in both channel and spatial dimensions. Jiang \textit{et al.} \cite{jiang2023mlic} introduced a multi-reference context model to capture local spatial, global spatial, and channel-wise contextual information. 
Kim \textit{et al.} \cite{kim2024diversify} utilized the diverse local, regional, and global contexts. 
While expanding the diversity of contextual information can boost compression performance, it also significantly increases the complexity of existing approaches. Therefore, a more efficient method for capturing diverse context is required to achieve a better trade-off between compression performance and complexity.

%% file: sec/3_method.tex
\section{Proposed Methods}

\subsection{Overview}
We follow the typical learned image compression scheme in our method. 
The encoder applies an analysis transform $g_a$ to the input image $x \in \mathbb{R}^{3 \times H \times W}$, generating latent representations $y=g_a(x|\phi) \in \mathbb{R}^{C \times H \times W}$. The latents are then rounded to obtain $\hat{y}=Q(y)$. The quantized latents are then losslessly entropy-coded through an entropy model $q_{\hat{Y}}(\hat{y})$. Finally, the decoder recovers $\hat{y}$ and generate reconstruction $\hat{x}$ through a synthesis transform $\hat{x}=g_s(\hat{y}|\theta)$. The notations $\phi$ and $\theta$ are trainable parameters of the analysis transform and synthesis transform, respectively.

Following the previous study \cite{zhang2024ggm}, we model the distribution of $\hat{y}$ as a generalized Gaussian model $\mathcal{N}_{\beta}(\mu,\alpha)$ with the shape parameter $\beta$ fixed as 1.5. The entropy model outputs the mean and scale parameters of the latents. We jointly use a hyperprior module and the proposed hierarchical progressive context model (HPCM) to estimate the entropy parameters of $\hat{y}$. The hyperprior module extracts side information $z$ to capture redundancy in latents. The side information is obtained through a hyper-analysis transform $z = h_a(y|\phi_h)$, which is then quantized $\hat{z} = Q(z)$. The quantized side information is used to estimate the entropy parameters through a hyper-synthesis transform $h_s(\hat{z}|\theta_h)$. The notation $\phi_h$ and $\theta_h$ are trainable parameters of the hyper-analysis and hyper-synthesis transform, respectively. Our HPCM divides latents into several groups and codes each group sequentially.
The entropy parameters for latents to be coded in the $i_{th}$ coding step are obtained through 
\begin{equation}
\begin{aligned}
    \mu_{i}, \alpha_{i} &= \operatorname{HPCM}(\hat{y}_{<i}, h_s(\hat{z}|\theta_h))
\end{aligned}
\end{equation}
where $\hat{y}_{<i}$ denotes previously coded latents before $i_{th}$ coding step. 

Our HPCM employs a hierarchical coding schedule to sequentially model contextual dependencies among latents at multiple scales. First, we partition the latents into multi-scale sub-latents through a specialized sampling method. Then, we code each scale of sub-latents sequentially, starting from the smallest scale and progressing to the largest, gradually modeling dependencies from long-range to short-range. This hierarchical coding schedule is introduced in Sec. \ref{HDS}. 
Furthermore, to better exploit the diverse context information at different coding stages, we present a progressive context fusion mechanism. Specifically, we incorporate the context from previous coding steps into the current step's contextual representations through a cross-attention mechanism. This approach enables the progressive accumulation of diverse context information. This progressive context fusion is introduced in Sec. \ref{PCF}.

\begin{figure*}[!t]
    \centering
    \centerline{\includegraphics[width=0.95\textwidth]{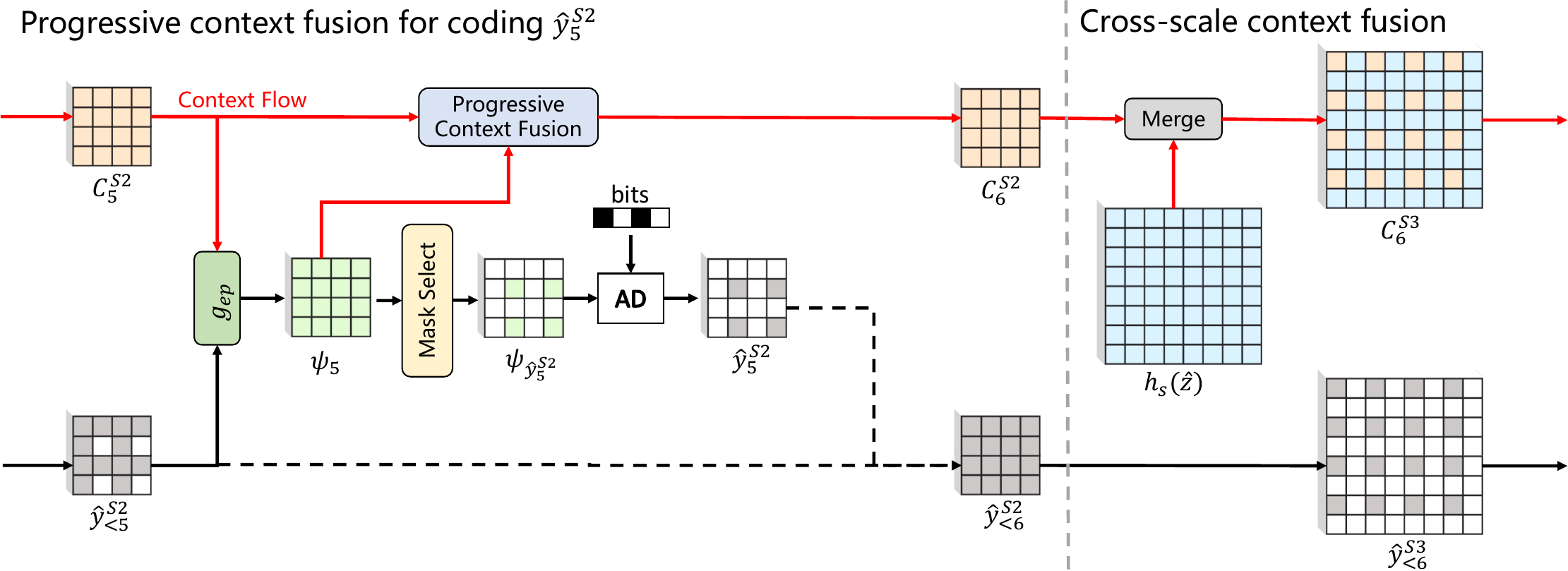}}
    \vspace{-0.5em}
    \caption{
    Diagram of our proposed Progressive Context Fusion (PCF) method. We take the coding process of $\hat{y}_{5}$ in $\hat{y}^{S2}$ and $\hat{y}_{6}$ in $\hat{y}^{S3}$ for example. $C_{i}^{S2}$, $\psi_{i}$ and $\psi_{\hat{y}_{i}}$ denote the progressively accumulated context at S2, entropy parameter state, and the entropy parameters for coding $\hat{y}_{i}$, respectively. $g_{ep}$ represents the entropy parameter network. 
    The detailed process of the PCF module is shown in Fig. \ref{attention}.
    In the cross-scale scenario, $C_{6}^{S2}$ and $h_s(\hat{z})$ are merged to obtain $C_{6}^{S3}$ at higher scale.
    }
    \vspace{-0.5em}
    \label{gather}
\end{figure*}

For end-to-end training, we optimize the model with the rate-distortion cost
\begin{equation}
\label{eq:rd}
    L=\mathcal{R}(\hat{y})+\mathcal{R}(\hat{z})+\lambda \cdot \mathcal D(x, \hat{x})
\end{equation}
where $\mathcal{R}(\hat{y})$ and $\mathcal{R}(\hat{z})$ denote the bitrates of $\hat{y}$ and $\hat{z}$; $\mathcal D(x, \hat{x})$ denotes the distortion between $x$ and $\hat{x}$; $\lambda$ controls the rate-distortion tradeoff.

\subsection{Hierarchical Coding Schedule}
\label{HDS}
Utilizing long-range context information is crucial for more accurate distribution estimation. 
Some learned image compression methods \cite{qian2022entroformer, li2024mixer} have exploited long-range context information in the latent domain through global attention mechanisms, demonstrating advanced compression performance. However, these methods also introduce higher complexity due to explicitly modeling long-range dependencies in the latent domain.
Our HPCM addresses this limitation through a hierarchical coding schedule.
As illustrated in Fig. \ref{overall_architecture}, we partition $\hat{y}$ into three sub-latents at different scales: $\hat{y}^{S1}$, $\hat{y}^{S2}$, and $\hat{y}^{S3}$. Coding smaller sub-latents first and then capturing context information on them enables a larger spatial receptive field, thereby facilitating more efficient long-range context modeling.
Additionally, following the joint spatial-channel context model \cite{li2023neural}, we apply different spatial partition strategies on eight evenly sliced channel groups. This allows for the interaction of spatial and channel-wise context information. 
The coding process of two different channel groups is shown in Fig. \ref{overall_architecture} to illustrate our hierarchical coding schedule.

We apply sequential conditional entropy coding at each scale and progressively exploit context information in each coding step. Specifically, $\hat{y}^{S1}$ is a 4× down-scaled version of $\hat{y}$. The first entropy coding step for $\hat{y}^{S1}$ is conditioned only on the hyperprior information. After completing the first coding step, subsequent steps are conditioned on our HPCM. Once $\hat{y}^{S1}$ is coded, we fill it back into its corresponding locations in $\hat{y}^{S2}$. Similarly, after $\hat{y}^{S2}$ has been coded, we fill it back into its corresponding locations in $\hat{y}^{S3}$. This up-scaling process, from $\hat{y}^{S1}$ to $\hat{y}^{S3}$, gradually models dependencies from long-range to short-range, ensuring comprehensive context representation across all scales.

\begin{figure}[!t]
    \centering
    \centerline{\includegraphics[width=0.7\linewidth]{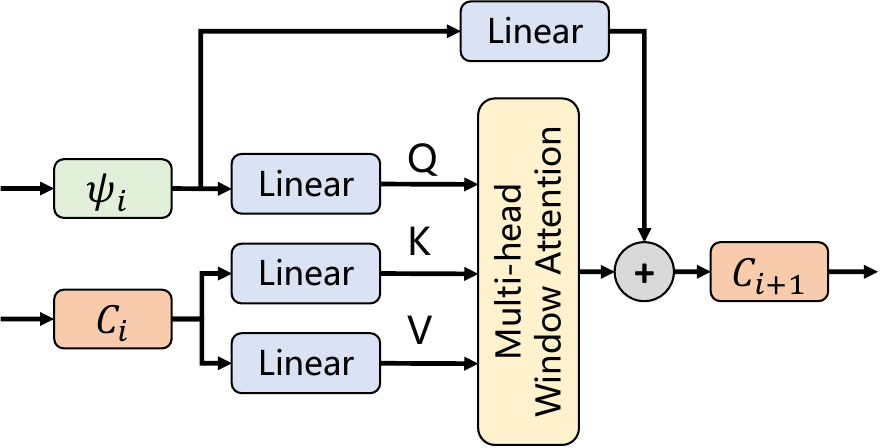}}
    \vspace{-0.5em}
    \caption{Illustration of Progressive Context Fusion (PCF) module. $C_{i}$ and $\psi_{i}$ denote the progressively accumulated context and entropy parameter state, respectively.}
    \vspace{-0.5em}
    \label{attention}
\end{figure}

To achieve a better balance between compression performance and complexity, we allocate different coding steps at each scale. Specifically, for $\hat{y}^{S1}$, the smallest scale, we employ an efficient 2-step coding process. This allows us to maintain low computational complexity while still capturing essential long-range context information. As the scale increases, the number of latents increases, requiring finer context modeling to save bits. We apply a 3-step quadtree-based entropy coding \cite{li2023neural} for $\hat{y}^{S2}$ and a 6-step octree-based entropy coding for $\hat{y}^{S3}$. We follow previous studies \cite{li2023flexible, mentzer2023m2t} to design the octree-based coding order at $\hat{y}^{S3}$. Specifically, within each 4×4 patch on the spatial dimension, we initially code latent elements widely dispersed across the spatial dimension. Subsequent coded latents are systematically positioned adjacent to previously coded latents. This approach exploits the spatial correlation in latents and enhances the overall context modeling ability. The detailed coding schedules are included in Sec. \ref{detailed_coding} of the supplement material.

\subsection{Progressive Context Fusion}
\label{PCF} 
Previous studies usually extract context reference from the hyperprior and the previously coded latents for entropy modeling. 
To better exploit the diverse contexts at different coding steps, our proposed progressive context fusion module incorporates the context information from previous coding steps into the current step's context reference. For the first coding step, the context information is down-scaled from the synthesized hyperprior. For the following coding steps, the context information $C_i$ is accumulated from previous coding steps. This approach aims to progressively accumulate diverse context information for more efficient utilization of context information for entropy coding. 

We take the coding for $\hat{y}_{i}^{S2}$ to illustrate the progressive fusion process. 
At $i_{th}$ coding step, the context information $C_{i}^{S2}$ is generated by fusing the context information $C_{i-1}^{S2}$ and the entropy parameter state $\psi_{i-1}$ at ${i-1}_{th}$ coding step. The entropy parameter state $\psi_{i-1}$ contains the entropy parameters used for coding $\hat{y}_{i-1}^{S2}$. As illustrated in Fig. \ref{gather}, the entropy parameter state $\psi_{i}$ in $i_{th}$ step is updated through
\begin{equation}
    \psi_{i}=g_{ep}(\operatorname{Concat}(\hat{y}_{<i}^{S2},C_{i}^{S2}))
\end{equation}
where $g_{ep}$ denotes the entropy parameter network. 
Then the corresponding entropy parameters $\psi_{\hat{y}_{i}^{S2}}$ for coding $\hat{y}_{i}^{S2}$ are selected from $\psi_{i}$. 
% The $\mu_{\hat{y}_{i}^{S2}}$ and $\alpha_{\hat{y}_{i}^{S2}}$ are further obtained by channel splitting of $\psi_{\hat{y}_{i}^{S2}}$. 
Similarly, $C_{i+1}^{S2}$ is obtained through combining $C_{i}^{S2}$ and $\psi_{i}$ for progressive context accumulation. As show in Fig. \ref{attention}, $C_{i}^{S2}$ and $\psi_{i}$ are fused through a cross-attention module. This fusion process can be formulated by
\begin{gather}
    Q = \operatorname{Linear}(\psi_{i}), \\
    K = \operatorname{Linear}(C_{i}^{S2}), \\
    V = \operatorname{Linear}(C_{i}^{S2}), \\
    C_{i+1}^{S2} = \operatorname{softmax}\left(\frac{QK^{T}}{\sqrt{d_{k}}}\right)V + \operatorname{Linear}(\psi_{i}),
\end{gather}
where $d_k$ is the dimension of $K$.

Additionally, to efficiently propagate the context information across multiple scales, we further propose a cross-scale context fusion method.
We take the context propagation process from $\hat{y}^{S2}$ to $\hat{y}^{S3}$ to illustrate this cross-scale context fusion.
As shown in Fig. \ref{gather}, to propagate the context information $C_{6}^{S2}$ at smaller scale to the context $C_{6}^{S3}$ at larger scale, we fill $C_{6}^{S2}$ to the corresponding locations in $C_{6}^{S3}$. For the remaining locations, we fill the synthesized hyperprior information $h_s(\hat{z})$. This approach efficiently combines the hyperprior information and cross-scale context information. 

Through this progressive context fusion at each coding step, the accumulated context $C_{i}$ progressively exhibits enhanced contextual diversity. 
Moreover, with the context fusion across scales, this progressively refined $C_{i}$ also enables efficient context fusion of both long-range and short-range dependencies in the hierarchical coding process, strengthening the entropy estimation capability.

\subsection{Other Improvements}
First, we conduct improvements in network structures in our model to strike a better balance between compression performance and complexity. Specifically, we utilized the advanced design of neural networks \cite{chollet2017xception, chen2023pconv, zhang2023practical, zhang2025learning} to construct transforms and context model networks. Additionally, we present a parameter-efficient way to construct context models, which shares the same network at different coding stages \cite{li2023neural}. This design significantly reduces model parameters while preserving essential context-aware adaptation capabilities. More details on the network structure are introduced in Sec.~\ref{overall_arch} and Sec.~\ref{structure_entropy} of the supplement material.
% 下面这个不确定要不要放在正文里，取决于最后表的部分怎么解读
Second, we optimized the entropy coder implementation through more efficient data exchange between Python and C, which enables efficient entropy coding.

%% file: sec/4_experiments.tex
\section{Experimental Results}

\begin{figure*}[!t]
    \centering
    \centerline{
        \begin{minipage}{0.45\linewidth}
         \includegraphics[width=\textwidth]{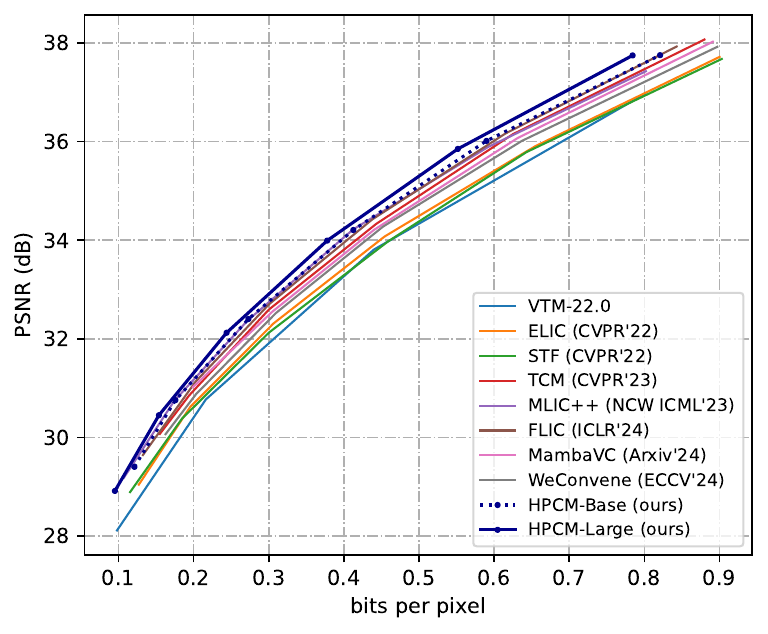}
         % \vspace{0.0001cm}
        \end{minipage}
        \hspace{1.2cm}
        \begin{minipage}{0.45\linewidth}
         \includegraphics[width=\textwidth]{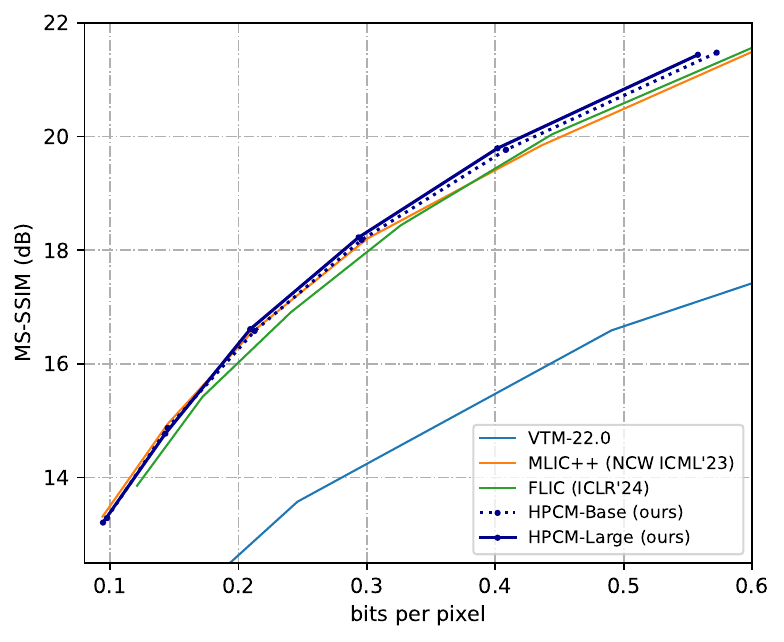}
         % \vspace{0.0001cm}
        \end{minipage}
    }
    \vspace{-1.0em}
    \caption{Rate-distortion curves on Kodak dataset.}
    \label{rd_kodak}
    \vspace{-1.0em}
\end{figure*}

\subsection{Experimental Settings}

\textbf{Training settings.} We use Flicker2W dataset \cite{liu2020unified} for training. During training, images are randomly cropped to 256 × 256 patches with a batch size of 32. The models are optimized by Eq. (\ref{eq:rd}). 
% We use both  and multi-scale structural similarity (MS-SSIM) \cite{wang2003multiscale} as distortion metrics. 
When optimized for mean square error (MSE) distortion metric, the Lagrange multiplier $\lambda$ belongs to $\{0.0018, 0.0035, 0.0067, 0.0130, 0.0250, 0.0483\}$. When optimized for multi-scale structural similarity (MS-SSIM) \cite{wang2003multiscale}, $\lambda$ belongs to $\{2.40, 4.58, 8.73, 16.64, 31.73, 60.50\}$. The Adam \cite{kingma2014adam} optimizer is used with $\beta_{1} = 0.9$ and $\beta_{2} = 0.999$. Our models are optimized with 2 million training steps. The learning rate is set as $10^{-4}$ initially, and reduced to $2\times 10^{-5}$ after 1.6M steps, then to $5\times 10^{-6}$ after 1.8M steps, and then to $10^{-6}$ after 1.9M steps.

\begin{figure}[!t]
    \centering
    \centerline{\includegraphics[width=0.45\textwidth]{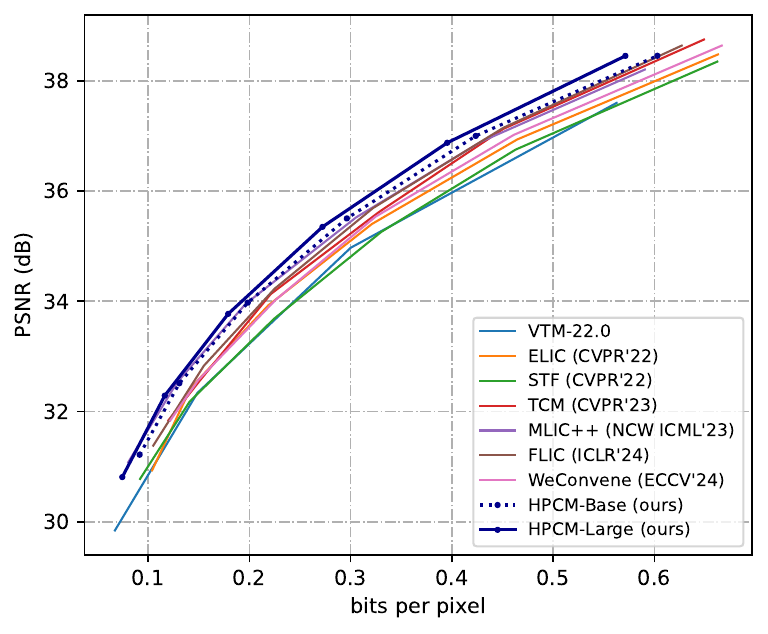}}
    \vspace{-1em}
    \caption{Rate-distortion curves on CLIC Pro Valid dataset.}
    \label{rd_clic}
    \vspace{-1em}
\end{figure}

\textbf{Evaluation settings.} We evaluate performance on three commonly used test datasets: Kodak dataset \cite{franzen1999kodak} which contains 24 images with 512 × 768 resolution; CLIC Professional Validation (CLIC Pro Valid) dataset \footnote{http://compression.cc} which contains 41 high-quality images; Tecnick dataset \cite{asuni2014testimages} which contains 100 images with 1200 × 1200 resolution. 
We use bits-per-pixel (bpp) to measure the bitrate and Peak Signal-to-Noise Ratio (PSNR) or MS-SSIM to measure the distortion.
We convert MS-SSIM to $-10\log_{10}(1-\text{MS-SSIM})$ for better comparison of rate-distortion curves. We also use the BD-Rate metric \cite{bjontegaard2001calculation} to evaluate rate saving. The anchor of BD-Rate in this paper is VTM-22.0\footnote{https://vcgit.hhi.fraunhofer.de/jvet/VVCSoftware\_VTM}. 
The coding time is evaluated on a single-core Intel(R) Xeon(R) Gold 6248R CPU and an NVIDIA GeForce RTX 3090 GPU. The model complexities, including kMACs/pixel and model parameter counts, are evaluated with DeepSpeed library\footnote{https://github.com/microsoft/DeepSpeed}.

\textbf{Model settings.} We present HPCM-Base and HPCM-Large targeting different complexities. The difference between the two models lies in the number of network layers in transform and entropy parameter network. The details are included in Sec. \ref{overall_arch} and Sec. \ref{structure_entropy} of the supplement material.

\subsection{Rate-Distortion Performance}
First, we verify the effectiveness of our proposed HPCM. Specifically, we train the CHARM \cite{minnen2020channel} and DCVC-DC intra \cite{li2023neural} entropy models using the same transform architecture as our HPCM-Base model. As shown in Table \ref{rd_time}, our proposed method achieves significantly higher compression performance, demonstrating the effectiveness of the proposed HPCM.

\begin{figure}[!t]
    \centering
    \centerline{\includegraphics[width=0.45\textwidth]{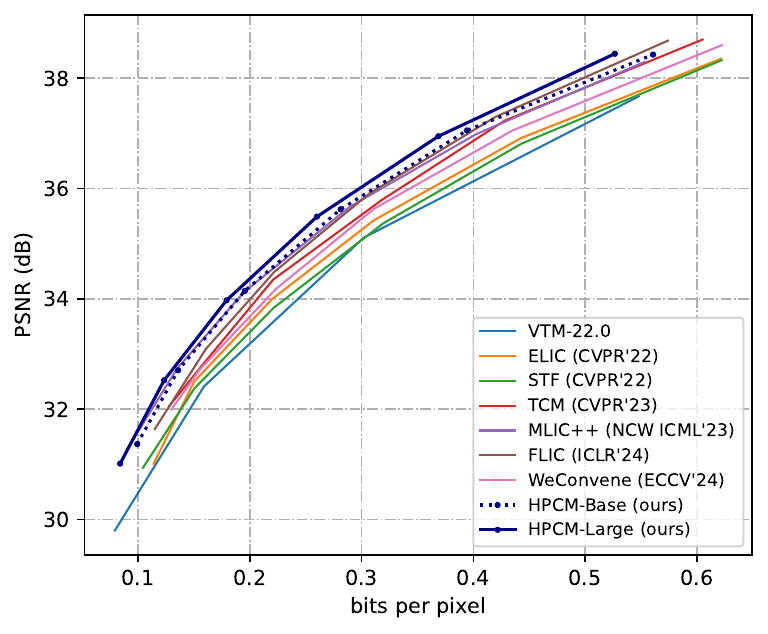}}
    \vspace{-1em}
    \caption{Rate-distortion curves on Tecnick dataset.}
    \label{rd_tecnick}
    \vspace{-1em}
\end{figure}

\begin{table*}[!t]
\centering
% \vspace{-0.8em}
\setlength{\tabcolsep}{3.5pt}
\caption{Compression performance and complexity comparison. VTM-22.0 is used as an anchor to calculate the PSNR BD-Rate. The best compression performance is marked in \textbf{bold}, and the second-best is \underline{underlined}.}
\vspace{-0.5em}
\begin{threeparttable}
\begin{tabular}{lccccccc}
\hline
\multirow{2}{*}{Model} & \multirow{2}{*}{\begin{tabular}[c]{@{}c@{}}Enc. Time$^\dagger$ \\ (ms)\end{tabular}} & \multirow{2}{*}{\begin{tabular}[c]{@{}c@{}}Dec. Time$^\dagger$ \\ (ms)\end{tabular}} & \multirow{2}{*}{\begin{tabular}[c]{@{}c@{}}kMACs\\ /pixel\end{tabular}} & \multirow{2}{*}{\begin{tabular}[c]{@{}c@{}}Params\\ (M)\end{tabular}} & \multicolumn{3}{c}{PSNR BD-Rate}                                                                                           \\ \cline{6-8} 
                       &                                                                            &                                                                            &                                                                         &                                                                       & Kodak                                 & CLIC Pro Valid                        & Tecnick                               \\ \hline
ELIC (CVPR'22) \cite{he2022elic}         & 126.5                                                                      & 111.4                                                                      & 573.88                                                                  & 36.93                                                                 & -3.22\%                               & -3.89\%                               & -4.57\%                               \\
STF (CVPR'22) \cite{zou2022devil}          & 142.5                                                                      & 156.8                                                                      & 511.17                                                                  & 99.86                                                                 & -2.06\%                               & 1.12\%                                & -2.17\%                               \\
TCM (CVPR'23) \cite{liu2023learned}          & 200.2                                                                      & 201.8                                                                      & 1823.58                                                                 & 76.57                                                                 & -10.70\%                              & -8.32\%                               & -11.84\%                              \\
MLIC++ (NCW ICML'23) \cite{jiang2023mlic}   & 193.4                                                                      & 226.4                                                                      & 1282.81                                                                 & 116.72                                                                & -15.15\%                              & -14.05\%                              & -17.90\%                              \\
FLIC (ICLR'24) \cite{li2023frequency}         & \textgreater{}1000                                                         & \textgreater{}1000                                                         & 1096.04                                                                 & 70.96                                                                 & -13.20\%                              & -9.88\%                               & -15.27\%                              \\
MambaVC (Arxiv'24) \cite{qin2024mambavc}     & 235.6                                                                          & 246.2                                                                          & 813.80                                                                   & 47.88                                                                 & -8.72\%                               & -                                     & -                                     \\
WeConvene (ECCV'24) \cite{fu2024weconv}    & 343.6                                                                      & 256.5                                                                      & 2343.13                                                                 & 107.15                                                                & -6.98\%                               & -5.66\%                               & -8.63\%                               \\ \hline
CHARM*                 & 57.5                                                                       & 70.6                                                                       & 495.75                                                                  & 58.53                                                                 & 0.86\%                                & 1.55\%                                & -1.32\%                               \\
DCVC-DC intra*         & 57.8                                                                       & 58.2                                                                       & 542.14                                                                  & 45.51                                                                 & -9.18\%                               & -8.54\%                               & -10.18\%                              \\ \hline
HPCM-Base (ours)       & 81.8                                                                       & 81.3                                                                       & 918.57                                                                  & 68.50                                                                  & \underline{-15.31\%} & \underline{-14.23\%} & \underline{-18.16\%} \\
HPCM-Large (ours)      & 91.2                                                                       & 90.2                                                                       & 1261.29                                                                 & 89.71                                                                 & \textbf{-19.19\%}                     & \textbf{-18.37\%}                     & \textbf{-22.20\%}                     \\ \hline
\end{tabular}
\begin{tablenotes}
\footnotesize
% \vspace{-0.25em}
\item $^{*}$The transforms are the same as our HPCM-Base model, and the entropy models are different.
% \item $^\dagger$ The coding time includes network inference and arithmetic coding time. A breakdown of these components is provided in Sec. 2 of the supplementary material.
% \vspace{-0.2em}
\item $^\dagger$ Coding time includes network inference time and arithmetic coding time. Details are presented in Sec. \ref{coding_time} of the supplementary material.
% \vspace{-1em}
\end{tablenotes}
\end{threeparttable}
\label{rd_time}
\vspace{-0.5em}
\end{table*}

\begin{table}[!t]
\centering
% \vspace{-0.8em}
\caption{Abaltion studies on hierarchical coding schedule.}
\vspace{-0.5em}
\begin{tabular}{ccc}
\hline
Model settings           & kMACs/pixel & BD-Rate \\ \hline
HPCM-Base*                & 918.57      & 0.00\%  \\
w/o hierarchical extraction & 1107.48     & 1.07\%  \\ \hline
coding step (2, 3, 3)    & 663.90      & 2.39\%  \\
coding step (2, 3, 12)   & 1427.91     & -2.55\% \\
coding step (4, 3, 6)    & 925.59      & 0.35\%  \\ \hline
\end{tabular}
\begin{tablenotes}
    \footnotesize
    \item $^{*}$ Our default setting is coding step (2, 3, 6).
\end{tablenotes}
\label{abla_hie}
\vspace{-0.5em}
\end{table}

Then, we compare our proposed two models, HPCM-Base and HPCM-Large, to state-of-the-art (SOTA) learned image compression methods, including \cite{he2022elic, zou2022devil, liu2023learned, jiang2023mlic, li2023frequency, qin2024mambavc, fu2024weconv}. 
As shown in Fig. \ref{rd_kodak}, our HPCM-Base model achieves better performance compared to other advanced methods on both PSNR and MS-SSIM metrics. Our HPCM-Large model achieves higher performance, showing about 0.2dB higher PSNR compared to other advanced methods at the same bitrate. Fig. \ref{rd_clic} and Fig. \ref{rd_tecnick} demonstrate our superior performance on the CLIC Pro Valid and Tecnick datasets. 
Table \ref{rd_time} shows the BD-Rate performance of our methods. Compared to VTM-22.0, our HPCM-Large model achieves 19.19\%, 18.37\%, and 22.20\% BD-Rate gain on Kodak, CLIC Pro Valid, and Tecnick datasets, respectively.

\subsection{Complexity}
Table~\ref{rd_time} shows the model complexity of our method and various SOTA learned image compression methods, including encoding and decoding times, kMACs/pixel, and model parameters. Benefiting from our efficient hierarchical progressive context modeling and improved network architecture, our method achieves a better trade-off between compression performance and complexity. 
Compared to MLIC++, our HPCM-Base model achieves comparable performance with much lower kMACs/pixel, and our HPCM-Large model achieves significantly higher performance with comparable kMACs/pixel.
As for the coding time, the total runtime reported in Table~\ref{rd_time} includes both network inference time and arithmetic coding time. A detailed breakdown of coding time is provided in Sec. \ref{coding_time} of the supplementary material.
% Benefiting from our optimized arithmetic coder, our models achieve faster coding times.

\subsection{Visual Comparison}
\begin{figure*}[!t]
    \centering
    \centerline{\includegraphics[width=1.00\textwidth]{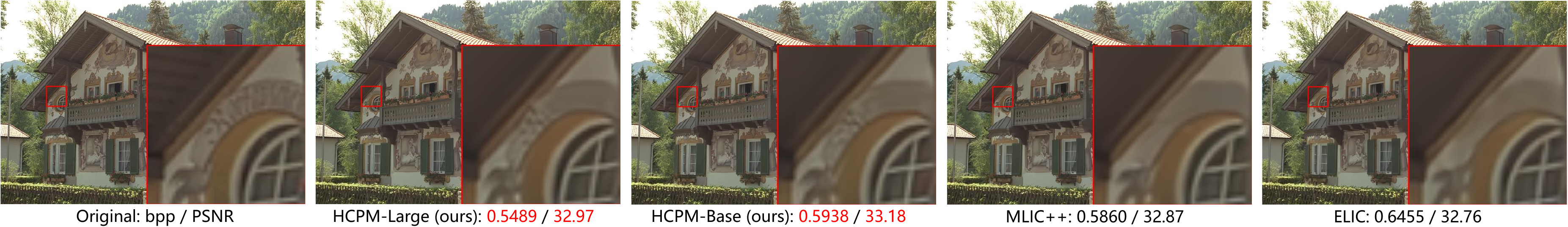}}
    \vspace{-0.8em}
    \caption{Visualization of reconstructed images of Kodim24 in the Kodak dataset with various learned image compression methods.
    }
    \label{visual}
    \vspace{-0.8em}
\end{figure*}

Figure \ref{visual} visualizes reconstructed images of Kodim24 in the Kodak dataset with different learned image compression methods. In some specific texture regions, our methods can keep more details.

\subsection{Ablation Studies}

\subsubsection{Ablation Studies on Hierarchical Coding Schedule}
To show the effectiveness of our hierarchical coding schedule, we conduct ablations on our HPCM-Base model.
First, we evaluate the model without the hierarchical context extraction, where the context information at different steps is exploited at the largest scale.
As shown in Table \ref{abla_hie}, this model results in higher computational complexity due to context modeling at the original latent scale. Additionally, since exploiting context information at a smaller scale enables more efficient long-range context modeling, our hierarchical context extraction achieves better performance.
More ablation studies on hierarchical coding stages are presented in Sec. \ref{add_abla_hie} of the supplementary material.

\begin{figure}[!t]
    \centering
    \centerline{\includegraphics[width=0.4\textwidth]{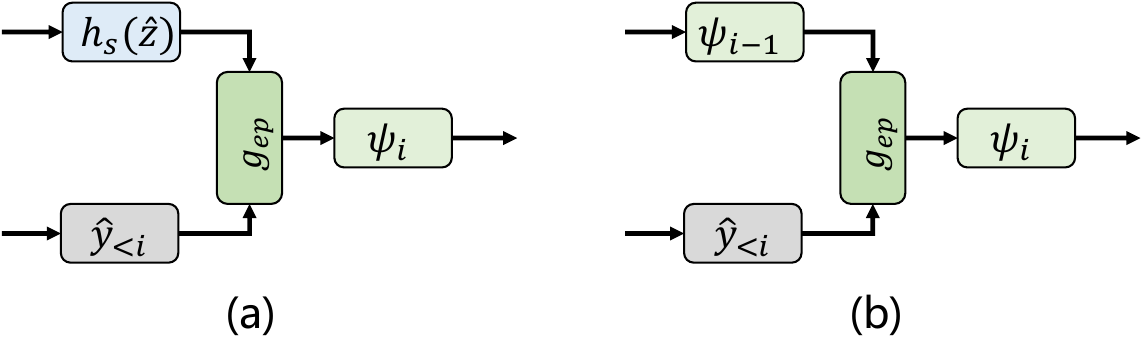}}
    \vspace{-0.5em}
    \caption{Illustration of methods to obtain the entropy parameter state $\psi_{i}$ in Table \ref{abla_pro}. (a) Without progressive fusion. (b) Use $\psi_{i}$ as progressive context.}
    \label{abl_pcf}
    \vspace{-0.5em}
\end{figure}

\begin{table}[!t]
\centering
\setlength{\tabcolsep}{3.5pt}
\caption{Abaltion studies on progressive context fusion.}
\vspace{-0.5em}
\begin{tabular}{ccc}
\hline
Model settings                                                  & kMACs/pixel & BD-Rate \\ \hline
HPCM-Base                                                        & 918.57      & 0.00\%  \\
w/o progressive fusion                                         & 872.80      & 4.71\%  \\
use $\psi_{i}$ as progressive context & 872.80      & 1.17\%  \\ \hline
\end{tabular}
\label{abla_pro}
\end{table}

We also show the influence of the number of coding steps allocated in different scales. In our HPCM model, the number of coding steps allocated for coding $(\hat{y}^{S1}, \hat{y}^{S2}, \hat{y}^{S3})$ is (2, 3, 6). Table \ref{abla_hie} shows the performance of other settings. 
The detailed coding schedules of these settings are introduced in Sec. \ref{detailed_coding} of the supplement material.
Since there are more latents at larger scales, the performance gain of using more coding steps at the largest scale is more significant. In contrast, at the smallest scale, increasing the number of coding steps has minimal influence. 
Although the model with coding step (2, 3, 12) further achieves 2.55\% rate saving over HPCM-Base, the computational complexity is much higher. To achieve a better balance between compression performance and complexity, we adopt the coding step as (2, 3, 6) in our methods.

\subsubsection{Ablation Studies on Progressive Context Fusion}
In this section, we show the effectiveness of our proposed progressive context fusion method. 
In the model without progressive context fusion, as shown in Fig.~\ref{abl_pcf} (a), we use the hyperprior $h_s(\hat{z})$ and previous coded latents $\hat{y}_{<i}$ to get $\psi_{i}$ through the entropy parameter network $g_{ep}$ at $i_{th}$ coding step. This approach is widely used in recent learned image compression methods. 
% As shown in Table~\ref{abla_pro}, there is a 4.71\% performance drop without progressive fusion, improving the effectiveness of our progressive context fusion method.
As shown in Table~\ref{abla_pro}, although the progressive fusion module introduces a slight increase in complexity, removing it results in a 4.71\% performance drop, demonstrating the effectiveness of our progressive context fusion method.

We also compare different ways to achieve progressive context fusion. As illustrated in Fig.~\ref{abl_pcf} (b), for comparison, we use $\psi_{i-1}$ and $\hat{y}_{<i}$ to obtain $\psi_{i}$ for entropy coding at $i_{th}$ step. Since the entropy parameter state $\psi_{i}$ retains context information from the previous $i-1$ coding steps, using $\psi_{i}$ as the progressive accumulated context achieves comparable performance. However, introducing $C_{i}$ as the progressive context information through a cross-attention mechanism yields better performance. While both methods leverage context information from previous coding steps, the cross-attention mechanism used with $C_i$ in our methods allows for more precise context integration, leading to improved performance.

\subsection{Visualization Analysis}

\begin{figure}[!t]
    \centering
    \centerline{\includegraphics[width=0.35\textwidth]{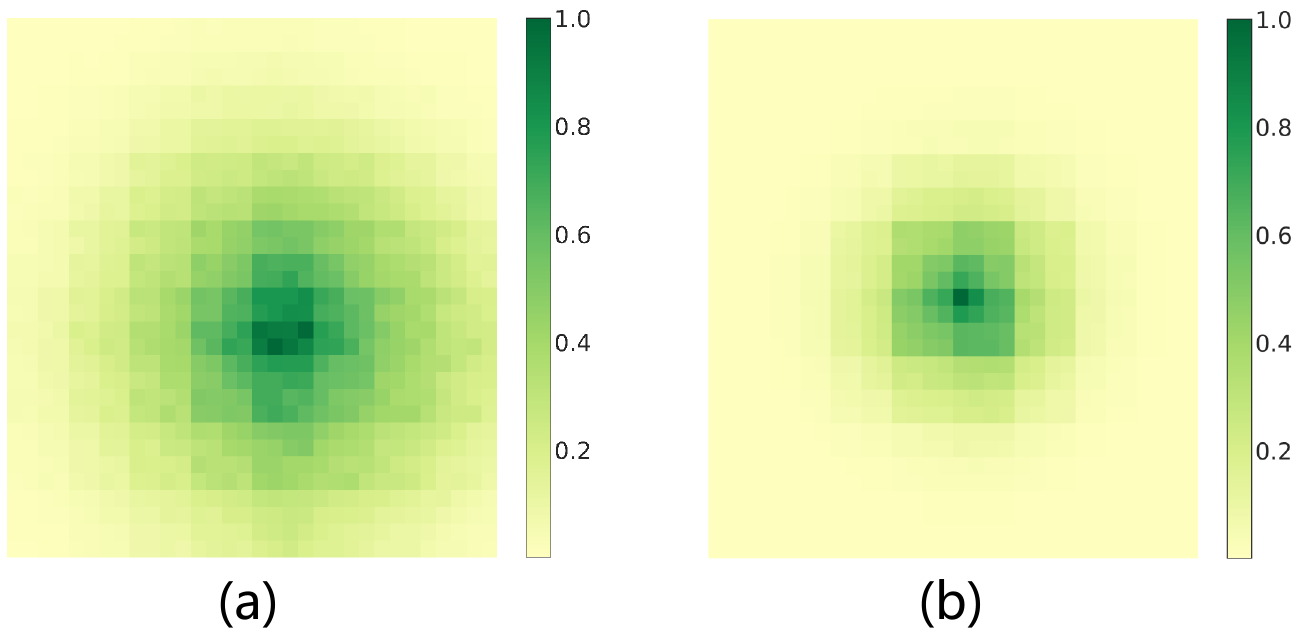}}
    \caption{Effective receptive fields (ERF) \cite{luo2016understanding} visualization of the entropy parameter network (a) $g_{ep}^{S1}$ and (b) $g_{ep}^{S3}$ in our HPCM-Base model. A wider dark area indicates a larger ERF.}
    \label{erf}
    \vspace{-1.0em}
\end{figure}

\textbf{Hierarchical Coding Schedule captures both long-range and short-range dependencies.}
Figure \ref{erf} visualizes the effective receptive field (ERF) \cite{luo2016understanding} of entropy parameter networks in our HPCM-Base model at different scales: $g_{ep}^{S1}$ and $g_{ep}^{S3}$. 
We visualize the gradient flow from the output of the entropy parameter networks to the input latents. The results are averaged on the Kodak dataset. The visualizations show that $g_{ep}^{S1}$ exhibits a relatively large ERF, while the ERF of $g_{ep}^{S3}$ is smaller, indicating that context extraction at $\hat{y}^{S1}$ captures long-range global patterns, while context extraction at $\hat{y}^{S3}$ focuses on short-range detailed patterns.

\begin{figure}[!t]
    \centering
    \centerline{\includegraphics[width=0.50\textwidth]{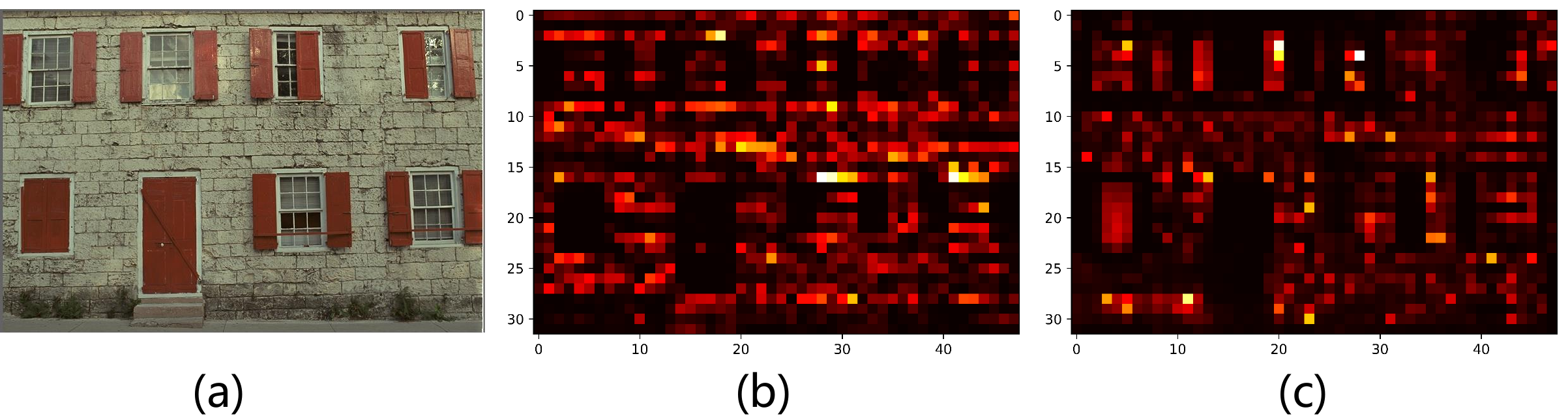}}
    \caption{Visualization of (a) Original image Kodim01 from the Kodak dataset; (b) Bit allocation map of latent $\hat{y}_{<6}^{S3}$. (c) Attention map of the progressive context fusion module at $6_{th}$ coding step on $\hat{y}^{S3}$.}
    \label{attn_map}
    \vspace{-1.0em}
\end{figure}

\textbf{Cross-attention in Progressive Context Fusion (PCF) module focuses more on high-bitrate regions.}
Figure ~\ref{attn_map} visualizes the bit allocation map of $\hat{y}_{<6}^{S3}$ and the attention map of the cross-attention operation in the PCF module at $6_{th}$ coding step on $\hat{y}^{S3}$ in our HPCM-Base model. For visualizing, we select the center point as the query in each window attention and rearrange attention maps to match the original latent size. 

The bit allocation map in Fig. \ref{attn_map} (b) shows that our model allocates higher bitrates to regions with more complex textures, which are more challenging to encode. The attention map in Fig. \ref{attn_map} (c) also assigns larger values to these high-bitrate regions. This indicates that our PCF module progressively leads $C_i$ to focus more on these high-bitrate regions that are more challenging for context modeling, thereby improving the accuracy of the entropy model.

%% file: sec/5_conclusion.tex
\section{Conclusion}

In this paper, we propose a novel hierarchical progressive context model for more efficient entropy coding in learned image compression. First, we introduce a hierarchical coding schedule to efficiently enable long-range context utilization. Furthermore, we propose a progressive context fusion module, which incorporates contextual information from previous coding steps and progressively accumulates the context information to efficiently enhance context diversity. Experimental results demonstrate that our proposed method achieves state-of-the-art performance and has a better trade-off between compression performance and complexity. 